\documentclass[journal]{IEEEtran}
\usepackage[cmex10]{amsmath}

\usepackage{cite}

\ifCLASSINFOpdf
   \usepackage[pdftex]{graphicx}
\else
\fi
 \usepackage{breqn}
\usepackage{epstopdf}

\usepackage{algorithmic}

\usepackage{array}

\usepackage{mdwmath}
\usepackage{mdwtab}
\usepackage{eurosym}
\usepackage{multicol}
\usepackage{eqparbox}
\usepackage{pdflscape}
\usepackage{pdfpages}
\usepackage{caption}
\usepackage{subcaption}
\usepackage{enumitem}

\begin{document}
%
\title{Boundaries of Operation for Refurbished Parallel AC-DC Reconfigurable Links in Distribution Grids}
%
%
%

\author{Aditya Shekhar, 
         Laura~Ram\'{i}rez-Elizondo and
        Pavol~Bauer
}

%
%

\markboth{This preprint is intended to be submitted for review as a future Journal}%
{Shekhar \MakeLowercase{\textit{et al.}}: }
%



\maketitle

\begin{abstract}
Parallel ac-dc reconfigurable link technology can find interesting applications in medium voltage power distribution. A given system can operate in different configurations while maintaining equivalent capacity during (n-1) contingencies. It is proved that within the defined operating boundaries, a parallel ac-dc configuration has higher efficiency as compared to pure ac or pure dc power delivery. Using sensitivity analysis, the variations in these efficiency boundaries with power demand, power factor, grid voltages, link lengths, conductor areas and converter efficiency is described. It is shown that parallel ac-dc system can have smaller payback time as compared to a purely dc power transmission for the same capacity due to lower investment cost in converter station and superior efficiency. As compared to a purely ac system, the payback of a refurbished parallel ac-dc configuration can be less than 5 years for a 10\,km, 10\,kV distribution link within the specified assumptions and operating conditions.
\end{abstract}


%
\IEEEpeerreviewmaketitle
\section{Introduction}
Increasing interest in the potential applications of medium voltage dc (MVDC) distribution is due to the opportunities it presents for flexible power redirection, grid reinforcement and decongestion~\cite{siemens}. An impending challenge faced by the distribution network operators (DNOs) is the issue of power delivery capacity enhancement of the existing ac grid infrastructure, wherein restructuring the infrastructure using flexible dc links could offer advantages~\cite{aditaylor,remus1}. 
\subsection{Converting Medium Voltage AC Distribution Line to DC}
\begin{figure}[!h]
\centering
\includegraphics[width=1\columnwidth]{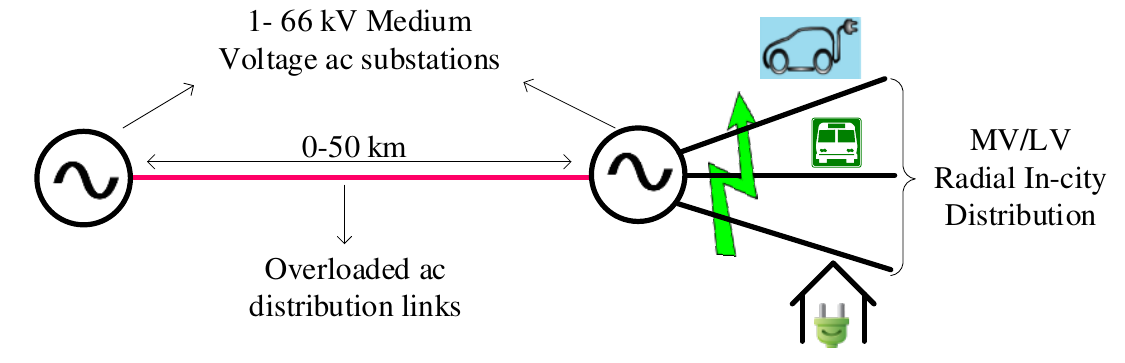}
\caption{Critical medium voltage ac distribution link overloaded due to increase in demand at the downstream radial network~\cite{adijepe}.}
\label{figbenchmark}
\end{figure}

The concept of converting the existing ac links for refurbished dc operation to maximize the power transfer capacity has recently gained significant attention~\cite{adijepe,ying,zhang,larru}. An illustration of capacity deficit benchmark ac distribution network is shown in~Fig.~\ref{figbenchmark}. 

It was shown in~\cite{adijepe} that the  total capacity of underground ac cables linking such a system can be improved by at least 1.5 times if they are refurbished to operate under dc conditions. The basis for this capacity increase was due to higher operating voltage (41\,\%), better thermal performance (1-3\,\%), better voltage regulation (3-5\,\%) and power factor correction (5-10\,\%). Similar or higher enhancement is possible if the system involves overhead lines~\cite{clerci,larru}. Superior efficiency of the employed ac/dc power electronic devices ensured that some efficiency gains over the original system are also achievable. Due to modularity, scalability, better harmonic performance and higher efficiency,  half bridge  ac/dc modular multilevel converters (MMC)~\cite{mmc0,mmc1,minos} were chosen for the proposed high power medium voltage dc link operation. 

The empirical proof of better insulation performance under imposed dc voltage as compared to ac is offered in~\cite{adienergies}. Studying the partial discharge behaviour in artificial voids of cable insulation under dc and different ac frequencies, the paper suggested that the refurbished dc cable link operating voltage ($V_{\text{dc}}$) at the peak of ac voltage ($\sqrt{2} v_{\text{ph,rms}}$) can be the conservative first assumption. The field implementation of renovated dc operation for capacity upgradation of XLPE cable line showed four years of uninterrupted operation with economic and reliability benefits as compared to ac solution~\cite{ying}. The study anticipated significant increase in achievable capacity gains and recommended a gradual rise in the operating dc~link~voltage with growing experience about the system. With such a gradual voltage upgradation strategy, modularity of the dc link MMC could prove advantageous~\cite{adijepe}.
\subsection{Reconfigurability with Parallel AC-DC Link Architecture}
The contingency analysis presented in~\cite{adispec} identified that a refurbished dc system may not necessarily maintain the enhanced capacity over the benchmark ac system in case certain faults occurred in the system. This realization led to the development of a reconfigurable parallel ac-dc link architecture that can maintain the enhanced power delivery capacity during (n-1) contingencies~\cite{adispec,adispeedam}. Specifically in~\cite{adispeedam}, while discussing the possible trade-offs in operating the system in different configurations, it was highlighted that a parallel ac-dc architecture could offer additional reliability, efficiency and cost benefits over completely ac or dc operation.

Among the earliest accounts of parallel operation of ac and dc lines for high voltage power transmission is found in~\cite{peterson}. In this work, greater flexibility of dc transmission as well as rapid control were considered advantageous for the composite system operation. Considering the effective conductor utilization with dc operation, in addition to the stability benefits, simultaneous operation of ac and dc in the same link is proposed in~\cite{rehman}. However, in the current work, parallel operation is referred to a system with
some conductors operating as ac links, while others as dc. Some challenges of parallel ac-dc system were explored in~\cite{reeve}, showing how the presence of an ac line introduces a coupling between the  front and back end dc link converter ac buses and increases the short circuit ratio. Over the years, the power system apparatus, corresponding device capabilities as well as grid requirements have significantly changed~\cite{dragan}. Consequently, the application of parallel ac-dc power delivery concept has broadened. Recognizing the advantage of using dc alongside ac distribution, authors in~\cite{acdclf} proposed a unified load flow model applicable to hybrid radial networks with distributed resources. The optimal operation of hybrid ac-dc power distribution is explored in~\cite{claudio,rodrigo}. An interesting recent research aims at interconnecting Crete island to Greek mainland using parallel ac-dc links~\cite{crete1,crete2}. The project explored the ability of these interconnectors for ensuring the continuity of supply, while minimizing the use of local generation in maintaining the capacity during (n-1) contingency.  In such circumstances, reconfigurable parallel ac-dc links could possibly improve the system availability~\cite{adispec,adispeedam}. 
\subsection{Research Focus and Key Contributions}
The focus of this paper is to prove that reconfigurable parallel ac-dc links are preferable, particularly for short distance (0-50\,km) particularly for high power (few tens of MVA), medium voltage levels (one to few tens of kV). The following key contributions are addressed in the paper:
\begin{itemize}
    \item Developing a generalized approach describing the possible system configurations and their capacity constraints.
    \item Establishing the equations describing the system losses under different configurations.
    \item Defining the efficiency boundaries for a case-study with varying link lengths, power demand and power factor.
    \item Performing the sensitivity analysis to describe the variation in the defined boundaries with grid voltage, link conductor area and average efficiency of station converter.
    \item Presenting an economic analysis comparing the payback of added investment associated with the proposed method.
\end{itemize}
\section{Generalized Description of Reconfigurable Parallel~AC-DC~Link System}
\label{sectwo}
\subsection{System Description}
\begin{figure}[!h]
\centering
\includegraphics[width=1\columnwidth]{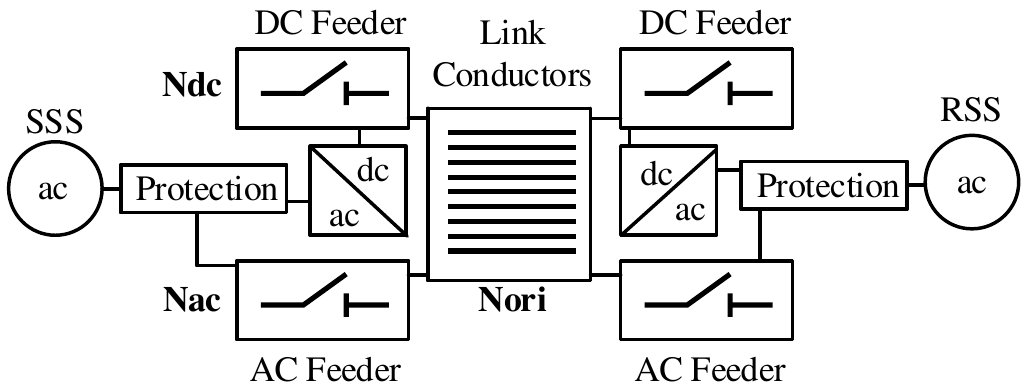}
\caption{Refurbished parallel ac-dc reconfigurable links in distribution grids.}
\label{figschematic}
\end{figure}

Let us consider that the sending end substation (SSS) and the receiving end substation (RSS) of an overloaded benchmark ac link system (as shown in~Fig.~\ref{figbenchmark}) are interconnected using a specified number of conductors (equal to $\text{N}_{\text{ori}}$). The schematic of the refurbished system with parallel ac-dc reconfigurable links is shown in~Fig~\ref{figschematic}. 

The power flow between SSS and RSS is directed by  the reconfiguration of switch blocks marked `DC Feeder' and `AC Feeder' that select the number of lines operating in dc and ac mode respectively. Depending on the combination of normally open (NO) and normally closed (NC) switches in these feeders, different system configurations are possible~(see~Section~\ref{secconfigurations}).

The link conductors can be either overhead lines or underground cables. Further, the cables can be single-cored or three-cored. The protection block at the ac side of each substation can be designed such that each ac or dc link acts as a point to point connection between SSS and RSS that can be isolated from ac side individually during faults. Alternatively, a common ac circuit breaker can be employed or dc breakers can also be explored~\cite{dcbreaker}. Similarly, design choices are available for the ac/dc converter blocks used for the dc link operation. For example, a single full capacity MMC can operate at each substation or the required capacity can be fulfilled using multiple partially rated MMCs for greater redundancy. A detailed discussion on different architectures, various choices available and the associated trade-offs in terms of reliability, efficiency and cost is presented in~\cite{adispec,adispeedam}.  In this paper, the results are shown with assumptions corresponding to single-cored underground cable with multiple MMCs per substation, each with rated capacity corresponding to a single dc link (see~Section~\ref{secconstraints}).
\subsection{Configuration Strategies}
\label{secconfigurations}
The pre-refurbished system (Fig.~\ref{figbenchmark}) originally operates as multiple 3-phase ac links, therefore, $\text{N}_{\text{ori}}$ is a multiple of three. In typical power systems, in consideration to the power constraints and redundancy requirements, $N_{\text{ori}}$ is most commonly 6 or 9. Depending on the state of the NO and NC switches in the feeders, different combinations of number of conductors operating in ac ($\text{N}_{\text{ac}}$) and dc ($\text{N}_{\text{dc}}$) are possible during normal operation.  $\text{N}_{\text{ac}}$ is always a multiple of three for three phase ac implementation, while $\text{N}_{\text{dc}}$ is even for a symmetric monopolar dc link implementation~\cite{minos_topology}. With the goal of maximizing the link conductor utilization $U_{\text{cond}}$, the refurbished ac-dc system can operate in $\left(\frac{N_{\text{ori}}}{3}\right)$ possible configurations while maintaining the desired capacity. In general, for any configuration `$C_{\text{n}}$', the number of conductors operating in ac mode ($N_{\text{ac,Cn}}$) is given by~\eqref{eqnac}.
\begin{align}
    N_{\text{ac,Cn}}&= (n-1)*\left( \frac{N_{\text{ori}}}{3}\right) \label{eqnac}
\end{align}
Where, n can take integer values between 1 to $\frac{N_{\text{ori}}}{3}$.  Correspondingly, the number of conductors operating in dc mode ($N_{\text{dc,Cn}}$) is given by~\eqref{eqndc}. 
\begin{equation}
   N_{\text{dc,Cn}}=
    \begin{cases}
     N_{\text{ori}}-N_{\text{ac,Cn}}, & \text{if}\ N_{\text{ori}}-N_{\text{ac,Cn}} \text{ is even} \\
      N_{\text{ori}}-N_{\text{ac,Cn}}-1, & \text{otherwise}
    \end{cases}\label{eqndc} \\
  \end{equation}  
  
The number of redundant conductors ($\text{N}_{\text{red}}$) during normal operation for any configuration is given by~\eqref{eqredundant},
\begin{align} 
    \text{N}_{\text{red}}&= \text{N}_{\text{ori}}-( \text{N}_{\text{ac,Cn}}+ \text{N}_{\text{dc,Cn}}) \label{eqredundant}
\end{align}
An extra configuration, referred to as `$C_{\text{0}}$' in this paper, corresponds to the conventional solution where additional conductors are installed by the DNOs for capacity enhancement in the ac link system. Based on the results presented in~\cite{adijepe}, $C_{\text{0}}$ needs 1.5 times more conductor area than the configuration with complete dc operation for equivalent capacity during (n-1) contingencies. Therefore, the total number of conductors $N_{\text{ac,C0}}$ is 
given by~\eqref{eqnacc0}.
\begin{align}
    N_{\text{ac,C0}}&=1.5*N_{\text{dc,Cn}}, \text{when,  } N_{\text{ac,Cn}}=0 \label{eqnacc0}
\end{align}

An example of possible configurations and corresponding number of conductors for $\text{N}_{\text{ori}}=9$ is shown in Table~\ref{tabone}.
\begin{table}[!h]
\renewcommand{\arraystretch}{1.2}
\caption{Number of ac, dc \& redundant conductors under operation for different configuration strategies with $\text{N}_{\text{ori}}=9$.}
\centering
\begin{tabular}{|c|c|c|c|c|}
\hline
 & $C_{\text{0}}$ &  $C_{\text{1}}$ & $C_{\text{2}}$ & $C_{\text{3}}$  \\  \hline
$\text{N}_{\text{ac}}$ & 12 & 0 & 3 & 6   \\  \hline
$\text{N}_{\text{dc}}$ & 0 & 8 & 6 & 2  \\  \hline
$\text{N}_{\text{red}}$ & 0 & 1 & 0 & 1 \\ \hline
\end{tabular} \label{tabone}
\end{table}

It can be inferred that for a given $N_{\text{ori}}$,  there are $\left(\frac{N_{\text{ori}}}{3}+1\right)$ possible configurations that have 50\,\% higher capacity than the benchmark system illustrated in~Fig.~\ref{figbenchmark}. These configurations have different efficiency boundaries for varying link length, power demand, RSS power factor (pf), grid voltage, converter efficiency and conductor area. Furthermore, each configuration has different investment costs and thus corresponding payback time. This paper provides insight on the highlighted aspects using the example of $N_{\text{ori}}=9$ based on the generalized equations that can be applied with alternative assumptions.
\subsection{Power delivery capacity constraints}
\label{secconstraints}
For a given configuration $C_{\text{n}}$, there are $\frac{\text{N}_{\text{ac,Cn}}}{3}$ three phase ac links and $\frac{\text{N}_{\text{dc,Cn}}}{2}$ dc links. The associated maximum power transfer capacities, $S_{\text{max,ac}}$ and $S_{\text{max,dc}}$, are given by~\eqref{eqsmaxac} and~\eqref{eqsmaxdc} respectively.
\begin{align}
    S_{\text{max,ac}}&=\frac{\text{N}_{\text{ac,Cn}}}{3}\cdot S_{\text{link}} \label{eqsmaxac} \\
    S_{\text{max,dc}}&=\frac{\text{N}_{\text{dc,Cn}}}{2}\cdot S_{\text{link}} \label{eqsmaxdc} 
\end{align}

Where, $S_{\text{link}}$ is the capacity of an individual link in the system, which can be assumed equal for a three conductor ac link and a two conductor dc link~\cite{adijepe}. 

The configurations employing parallel ac-dc links have an additional degree of freedom in controlling the active power flow using the dc link converters. The ratio of active dc power flow to that of ac is defined as 'y' and this can influence the efficiency of the system in meeting the given apparent power demand at the RSS ($S_{\text{actual}}$). However, y can be varied within certain limits, where, the minimum $ y_{\text{min}}$ and maximum $ y_{\text{max}}$ are given by~\eqref{eqymin} and~\eqref{eqymax}, respectively.
\begin{equation}
    y_{\text{min}}=
    \begin{cases}
      0, & \text{if}\ S_{\text{actual}}\leq S_{\text{max,ac}} \\
      \frac{S_{\text{actual}}-S_{\text{max,ac}}}{S_{\text{actual}}}, & \text{otherwise}
    \end{cases}\label{eqymin} \\
  \end{equation}
  \begin{equation}
    y_{\text{max}}=
    \begin{cases}
      1, & \text{if}\ S_{\text{actual}}\leq S_{\text{max,dc}} \\
      \frac{S_{\text{max,dc}}}{S_{\text{actual}}}, & \text{otherwise} \label{eqymax}
    \end{cases}
  \end{equation}

The precise value of these limits vary with pf, however this will not alter the efficiency boundaries under the operating conditions explored in this paper. Therefore, the validity of~\eqref{eqymin} and~\eqref{eqymax} is under the assumption of unity power factor (upf).
\subsection{Conductor temperature and resistance}
A range of ac and dc conductor currents, temperatures and resistances are possible for meeting the same apparent power demand, which influence the system level efficiency of each configuration. For a particular conductor current $I_{\text{cond}}$, the operating temperature $T_{\text{cond,k}}$ and conductor resistance $R_{\text{cond,k}}$ at $\text{k}^{th}$ iteration is computed based on \eqref{eqtcondk} and \eqref{eqrcondk} respectively. 
\begin{align}
    T_{\text{cond,k}}&= T_{\text{amb}}+\left(\frac{I_{\text{cond}}^{2}R_{\text{cond,k-1}}}{I_{\text{rated}}^{2}R_{\text{90$^{\circ}$C}}}\right)\left(90-T_{\text{amb}}\right) \label{eqtcondk} \\
    R_{\text{cond,k}}&=R_{\text{90$^{\circ}$C}}\left(\frac{1+\alpha(T_{\text{cond,k}}-T_{\text{amb}})}{1+\alpha(90-T_{\text{amb}})}\right) \label{eqrcondk}
\end{align}
Herein, $T_{\text{amb}}$ is the ambient temperature, $R_{\text{90$^{\circ}$C}}$ is the resistance of the conductor in $\Omega/km$ at 90$^{\circ}$C and $\alpha$ is the temperature coefficient. $R_{\text{Tcond,0}}$ is the initial conductor resistance assumed equal to $R_{\text{90C}}$. The iterative process terminates once the difference in the estimated resistance between two iterations becomes less than $1e^{-6}$. 
\subsection{Generic Equations Describing System Losses}
For any $N_{\text{ori}}$ the system loss equations can take three possible structure corresponding to 1) complete ac configuration  ($C_{0}$) 2) complete dc configuration ($C_{1}$) 3) parallel ac-dc configurations ($C_{n}$ with $n\neq (0,1)$). 
\subsubsection{Complete ac (Configuration $C_{0}$)} The  current per conductor  ($I_{\text{C0}}$) is given by~\eqref{eqiac},
\begin{align}
I_{\text{C0}} &= \frac{S_{\text{actual}}*10^3}{\left(\frac{N_{\text{ac,C0}}}{3}\right)\sqrt{3}V_{\text{LL,rms}}} =\frac{\sqrt{3}k}{(n_{\text{ori}}+3)} \label{eqiac} \\
\text{where,} & \nonumber \\
k &= \frac{S_{\text{actual}}*10^3}{V_{\text{LL,rms}}} \label{eqk}
\end{align}

$V_{\text{LL,rms}}$ is the line to line r.m.s. SSS bus voltage in kV and $S_{\text{actual}}$ is the RSS demand in MVA. The total link conductor loss $P_{\text{loss,C0}}$ is given by~\eqref{eqpac},
\begin{align}
P_{\text{loss,C0}} &= N_{\text{ac,C0}} I_{\text{C0}}^2 R_{\text{C0}} = \frac{3k^2L r_{\text{C0}}(T_{\text{C0}},A)}{n_{\text{ori}}+3} \label{eqpac}
\end{align}
Herein, L and A are the length and cross-sectional area of the cable conductor respectively. $r_{\text{C0}}(T_{\text{C0}},A)$ is the ohmic ac resistance per kilometer depending on the operating conductor temperature $T_{\text{C0}}$ and A. 
\subsubsection{Complete dc (Configuration $C_{1}$)} During normal operating conditions per conductor dc current $I_{\text{C1}}$ flows through $\left(\frac{n_{\text{dc,C1}}}{2}\right)$ dc links as given by \eqref{dcicon}
\begin{align}
I_{\text{C1}} &= \frac{S_{\text{actual}}\cos \theta*10^3}{\left(\frac{n_{\text{dc,C1}}}{2}\right)V_{\text{dc}}} =\sqrt{\frac{3}{2}}\left(\frac{k\cos \theta}{n_{\text{dc,C1}}}\right) \label{dcicon} \\
\text{where,} & \nonumber \\
V_{\text{dc}} &= \frac{2\sqrt{2}V_{\text{LL,rms}}}{\sqrt{3}}
\end{align}
$V_{\text{dc}}$ is chosen keeping in mind the insulation performance of the refurbished underground cables at $\sqrt{2}$ of the rms ac voltage rating~\cite{adienergies}. It is possible that the cable ac to dc voltage enhancement factor can be chose higher, leading to further benefits with refurbished dc operation, but this should be incrementally done with field trials~\cite{ying}.  $\cos \theta$ is the power factor of the load connected at the RSS. The total dc link conductor power loss $P_{\text{cond,C1}}$ is given by~\eqref{eqdcpowercon},
\begin{align}
P_{\text{cond,C1}} &= n_{\text{dc,C1}}I_{\text{C1}}^2 R_{\text{C1}} \nonumber \\
&= \frac{3k^2\cos^2 \theta L r_{\text{C1}}(T_{\text{C1}},A)}{2n_{\text{dc,C1}}} \label{eqdcpowercon}
\end{align}
Where, $r_{\text{C1}}(T_{\text{C1}},A)$ is the dc conductor resistance in ohm per km depending on operating temperature $T_{\text{C1}}$ and conductor area of cross-section. The converter loss $P_{\text{conv,C1}}$ is given by~\eqref{eqpconvdc1},
\begin{align}
P_{\text{conv,C1}} &= (1-\eta)S_{\text{actual}}\cos \theta*10^6 \nonumber \\
&= (1-\eta) kV_{\text{LL,rms}}\cos \theta *10^3 \label{eqpconvdc1}
\end{align}
$\eta$ is the efficiency of the MMC as computed in~\cite{adijepe,silvio}. From~\eqref{eqdcpowercon} and~\eqref{eqpconvdc1}, the total system losses $P_{\text{loss,C1}}$ are estimated using~\eqref{eqpsysdc},
\begin{align}
   P_{\text{loss,C1}} &= P_{\text{cond,C1}}+2P_{\text{conv,C1}} \label{eqpsysdc}
\end{align}
\subsubsection{Parallel ac-dc configurations ($C_{n}$ with $n\neq (0,1)$)} There are $\left(\frac{N_{\text{ori}}}{3}-1\right)$ possible parallel ac-dc configurations with structurally similar loss equations. The corresponding dc and ac conductor currents, $I_{\text{Cn,dc}}$ and $I_{\text{Cn,ac}}$, are given by~\eqref{eqidcdash} and~\eqref{eqiacdash}, respectively.
\begin{align}
I_{\text{Cn,dc}} &= \frac{yS_{\text{actual}}\cos \theta *10^3}{\left(\frac{n_{\text{dc,Cn}}}{2}\right)V_{\text{dc}}} =\sqrt{\frac{3}{2}} \frac{yk\cos \theta}{n_{\text{dc,Cn}}} \label{eqidcdash}\\
I_{\text{Cn,ac}} &= \frac{(1-y)S_{\text{actual}}\cos \theta*10^3}{\sqrt{3}\left(\frac{n_{\text{ac,Cn}}}{3}\right)V_{\text{LL,rms}}} =  \frac{\sqrt{3}(1-y)k\cos \theta}{n_{\text{ac,Cn}}} \label{eqiacdash} 
\end{align}
The power loss in the cable conductors is given by~\eqref{eqpacdc1},
\begin{multline}
P_{\text{cond,Cn}} = n_{\text{ac,Cn}}I^2_{\text{Cn,ac}}Lr_{\text{Cn,ac}}(T_{\text{Cn,ac}},A) \\ +n_{\text{dc,Cn}}I^2_{\text{Cn,dc}}Lr_{\text{Cn,dc}}(T_{\text{Cn,dc}},A)  \label{eqpacdc1}
\end{multline}
Substituting the value of conductor current from~\eqref{eqiacdash} and simplfying,
\begin{multline}
P_{\text{cond,Cn}}=3k^2L\cos^2\theta\left(\frac{(1-y)^2r_{\text{Cn,ac}}(T_{\text{Cn,ac}},A)}{n_{\text{ac,Cn}}} \right. \\ \left. {}\vphantom{} + \frac{y^2r_{\text{Cn,dc}}(T_{\text{Cn,dc}},A)}{2 \cdot n_{\text{dc,Cn}}} \right) \label{eqpacdc2}
\end{multline}
Here, $r_{\text{Cn,ac}}(T_{\text{Cn,ac}},A)$ and $r_{\text{Cn,dc}}(T_{\text{Cn,dc}},A)$ are the operating ac and dc resistances respectively, corresponding to temperatures $T_{\text{Cn,ac}}$ and $T_{\text{Cn,dc}}$, imposed by the operating currents $I_{\text{Cn,ac}}$ and $I_{\text{Cn,dc}}$. The converter loss $P_{\text{conv,Cn}}$ is given by~\eqref{eqpconvacdc}, 
\begin{align}
P_{\text{conv,Cn}} &= (1-\eta)\cdot y\cdot S_{\text{actual}}\cdot \cos \theta*10^6 \nonumber \\
&= (1-\eta)\cdot y\cdot k\cdot V_{\text{LL,rms}}\cdot \cos \theta *10^3 \label{eqpconvacdc}
\end{align}
It can be observed that these are lower than $P_{\text{conv,C1}}$ given in~\eqref{eqpconvdc1} by a factor of $y\leq 1$. The total system loss $P_{\text{loss,Cn}}$ in parallel ac-dc configuration can be estimated from conduction and converter losses similar to~\eqref{eqpsysdc}. Equations~\eqref{eqidcdash}-\eqref{eqpconvacdc} are valid for all parallel ac-dc configurations $C_{n}$ with $n\neq (0,1)$.
\section{Efficiency Boundaries For Different Configurations}
\label{secfive}
In this section the efficiency boundaries are derived for different configurations (C0-C3) corresponding to $N_{\text{ori}}=9$. The results are specific to $V_{\text{LL,rms}}=10$\,kV, aluminum conductors with $A=400$\,mm$^2$ and average converter efficiency of 99.34\,\% ~\cite{adijepe}. A sensitivity analysis showing the variation in the defined boundaries with these three parameters can be found in~Section~\ref{secsense}.
\subsection{System Losses and Crossover Points}
\begin{figure}[!h]
\centering
\includegraphics[width=1\columnwidth]{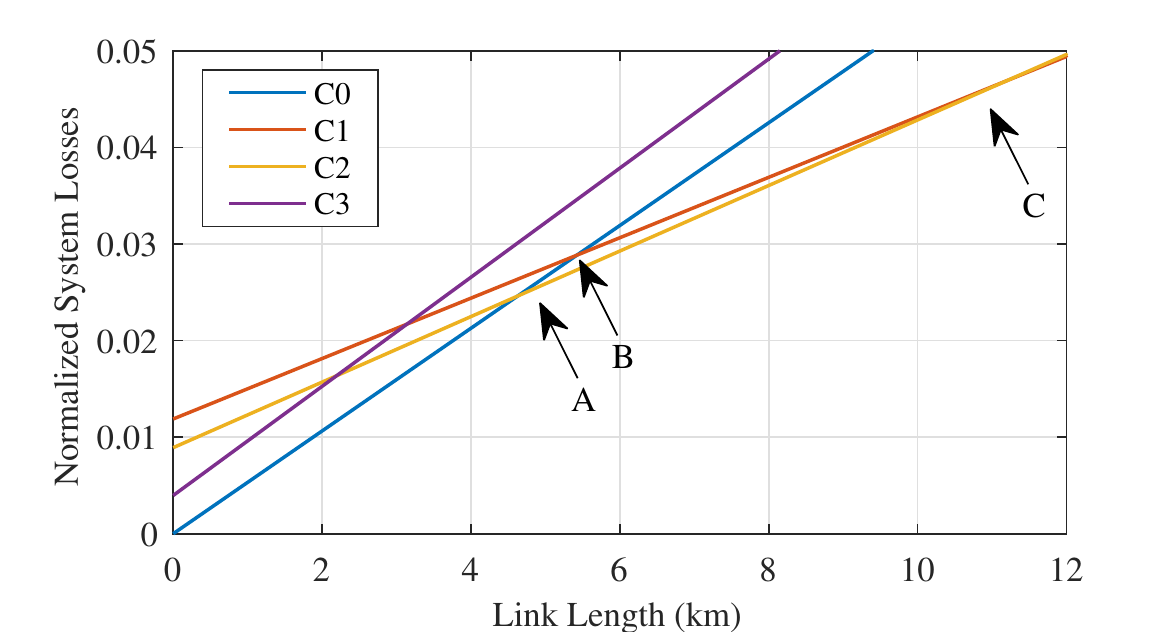}
\caption{Link length dependent normalized system power losses for different operational mode strategies.}
\label{figcrossovers}
\end{figure}

Based on the power loss equations derived in the previous section, the link length dependent system losses are shown in Fig.~\ref{figcrossovers} for different strategies C0-C3 delivering 3\,p.u $S_{\text{actual}}$ at 0.9 power factor. The depicted losses are normalized with respect to $S_{\text{actual}}$. The dc active power share y is considered 0.75 for C2 and 0.33 for C3 in this example, corresponding to the distribution factor at full load rated operation of the link system for the respective configuration. However, this value is not necessarily the optimal efficiency point and can be actively controlled within the limits defined by~\eqref{eqymin} and~\eqref{eqymax}.

It can be observed that configurations C1-C3 have higher losses as compared to C0 for lower link lengths, corresponding to the dc link converter losses. When a parallel ac-dc configuration such as C2/C3 is used, these losses are lower as a part of the power demand at RSS is delivered using the ac link. On the other hand, the link conductors losses in C0  dominate after crossover points A and B as compared to C2 and C1, respectively. The complete dc configuration C1 becomes more efficient than parallel AC-DC configuration C2 after crossover point C. C2 is the most efficient configuration for the link lengths between crossover points A and C, after which C1 should be considered. The crossover points A, B and C vary with $S_{\text{actual}}$, power factor, grid voltage, conductor area, converter efficiency and y of a particular system. Therefore, it is important to define these boundaries comprehensively to decide which operational mode and configuration to deliver the given power demand at maximum efficiency.
\subsection{Crossover Length for Different Strategies}
\begin{figure}[!h]
\centering
\includegraphics[width=1\columnwidth]{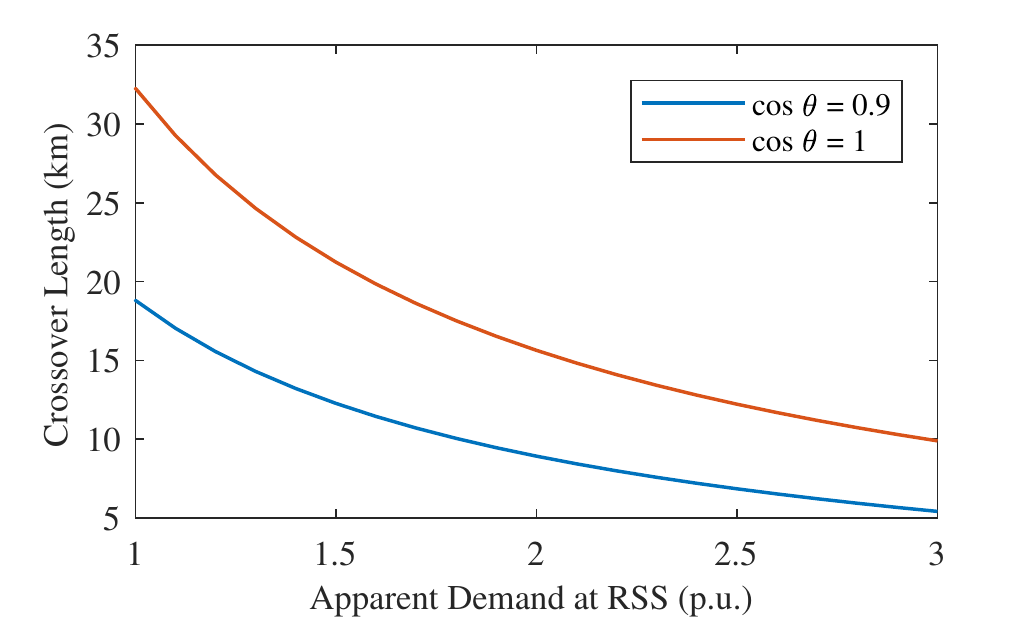}
\caption{Crossover for C0-C1 with respect to the normalized apparent demand ($S_{\text{actual}}$/$S_{\text{link}}$) for different pf (Point B).}
\label{figlcross_C0C1}
\end{figure}

The crossover length at point B ($L_{\text{cr,B}}$) can be derived by equating the system losses of C0 and C1  derived in~\eqref{eqpac} and~\eqref{eqpsysdc}. The derived expression is given by~\eqref{eqlcrit1},
\begin{align}
L_{\text{cr,B}} = \frac{2(1-\eta)\cos \theta}{3\left(\frac{r_{\text{C0}}}{n_{\text{ori}}+3} - \frac{ r_{\text{C1}}\cos^2 \theta}{2(n_{\text{ori}}-1)}\right)}\cdot \frac{V^2_{\text{LL,rms}}}{S_{\text{actual}}} \label{eqlcrit1}
\end{align}
Fig.~\ref{figlcross_C0C1} shows the crossover length $L_{\text{cr,B}}$ above which configuration C1 is more efficient for different $S_{\text{actual}}$ and pf.

\begin{figure}[!h]
\centering
\includegraphics[width=1\columnwidth]{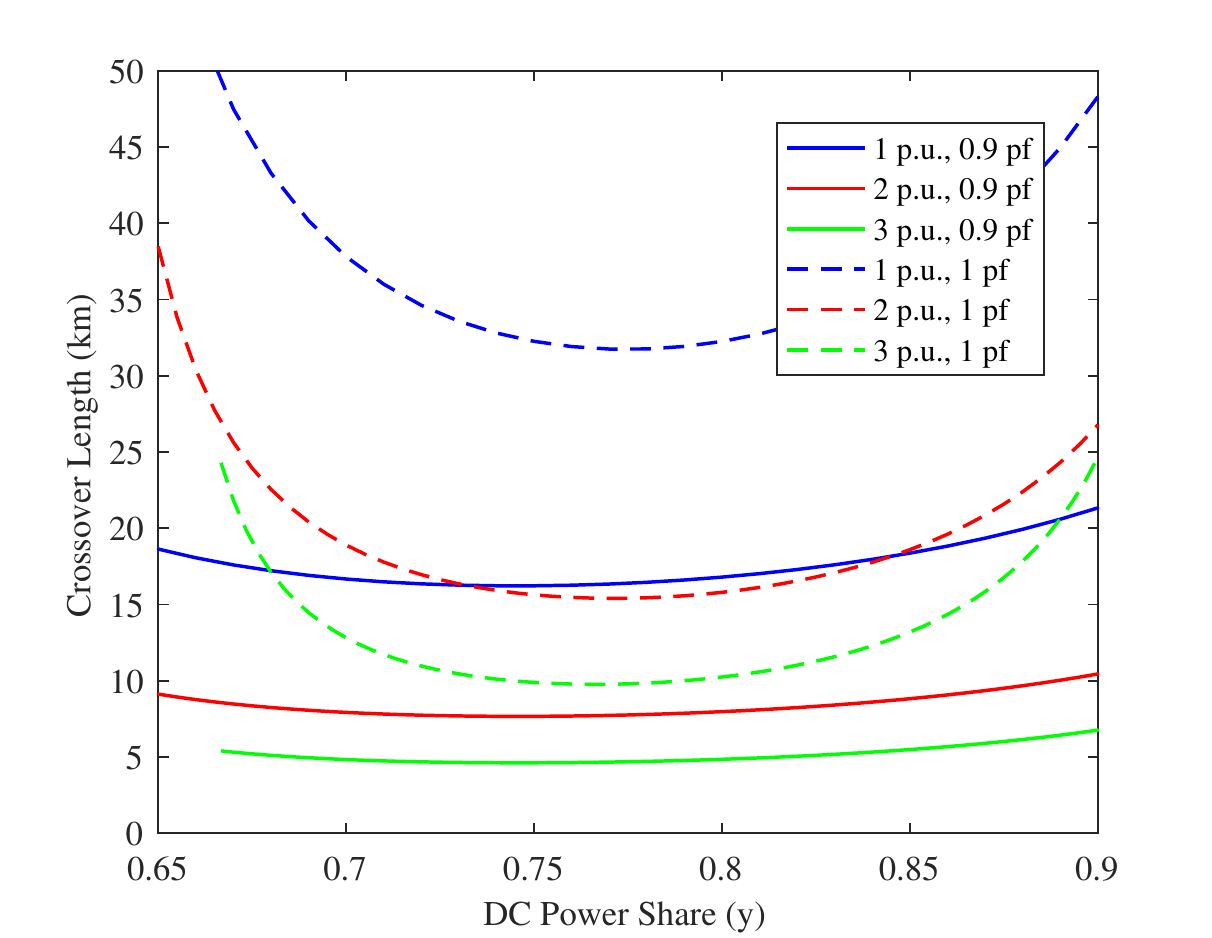}
\caption{Crossover for C0-C2 with respect to y for different normalized apparent demand ($S_{\text{actual}}$/$S_{\text{link}}$) and pf (Point A).}
\label{figlcross_C0C2}
\end{figure}

Similarly, $L_{\text{cr,A}}$ that describes the crossover link length above which the parallel ac-dc configuration C2 is operationally more efficient than C0, can be found by equating system losses of C0 and C2, as shown in~\eqref{eqlcrit2}.
\begin{align}
L_{\text{cr,A}} = \frac{2(1-\eta)y\cos \theta \left(\frac{V^2_{\text{LL,rms}}}{S_{\text{actual}}}\right)}{3\left(\frac{r_{\text{C0}}}{n_{\text{ori}}+3} - \cos^2 \theta \left(\frac{ (1-y)^2r_{\text{C2,ac}}}{\frac{n_{\text{ori}}}{3}} + \frac{ y^2r_{\text{C2,dc}}}{2\cdot \frac{2n_{\text{ori}}}{3}} \right)\right)} \label{eqlcrit2}
\end{align} 
 Since the dc link power can be actively controlled for a given $S_{\text{actual}}$, $L_{\text{cr,A}}$ varies with y as shown in Fig.~\ref{figlcross_C0C2}. When y is high, dc link converter losses are high, and when y is low, significant losses occur in the ac link of C2. In either case, the system is not operating at its optimal efficiency point. $y_{\text{opt}}$  varies with the link length, $S_{\text{actual}}$ and the pf and consequently, the minimum crossover length occurs at different values of y as can be observed.

\begin{figure}[!h]
\centering
\includegraphics[width=1\columnwidth]{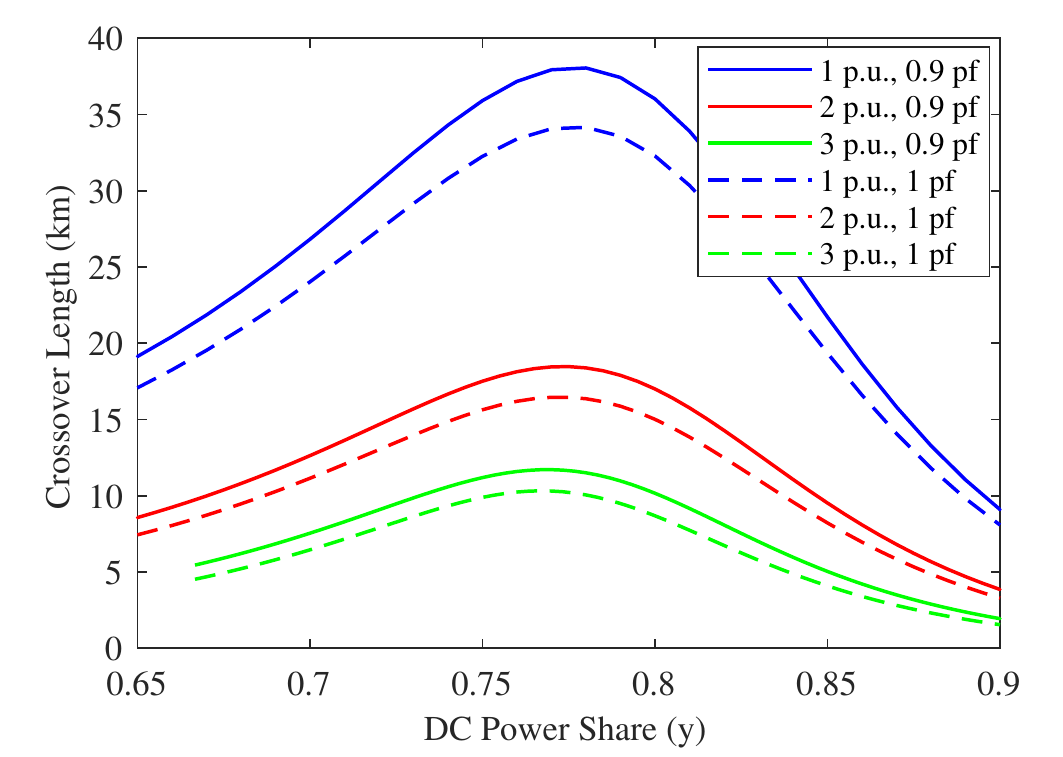}
\caption{Crossover length for C1-C2 with respect to the dc power share y for different apparent demand $S_{\text{actual}}$ and power factor (pf) at RSS (Point C).}
\label{figlcross_C1C2}
\end{figure}

By equating the system losses of C1 with C2, the crossover length $L_{\text{cr,C}}$ can be estimated from~\eqref{eqlcross},
\begin{align}
L_{\text{cr,C}} = \frac{2(1-\eta)(1-y) \left(\frac{V^2_{\text{LL,rms}}}{S_{\text{actual}}}\right)}{3\cos \theta\left(\frac{ (1-y)^2r_{\text{C2,ac}}}{\frac{n_{\text{ori}}}{3}} + \frac{ y^2r_{\text{C2,dc}}}{2\cdot \frac{2n_{\text{ori}}}{3}} -\frac{ r_{\text{C1}}}{2(n_{\text{ori}}-1)}\right)}
    \label{eqlcross}
\end{align}
The corresponding results for different $S_{\text{actual}}$, y and pf is shown in~Fig.~\ref{figlcross_C1C2}. As $S_{\text{actual}}$ increases, $L_{\text{cr,C}}$ decreases because dc link efficiency is greater than that of ac. Similarly, with increasing pf, the active power delivered with the link increases, giving lower $L_{\text{cr,C}}$. When y is either too low or too high, infrastructure utilization of C2 is not at its optimal point, resulting in lower $L_{\text{cr,C}}$. 
\subsection{Efficiency Boundaries}
\begin{figure}
\centering
\begin{subfigure}[b]{1\columnwidth}
   \includegraphics[width=1\linewidth]{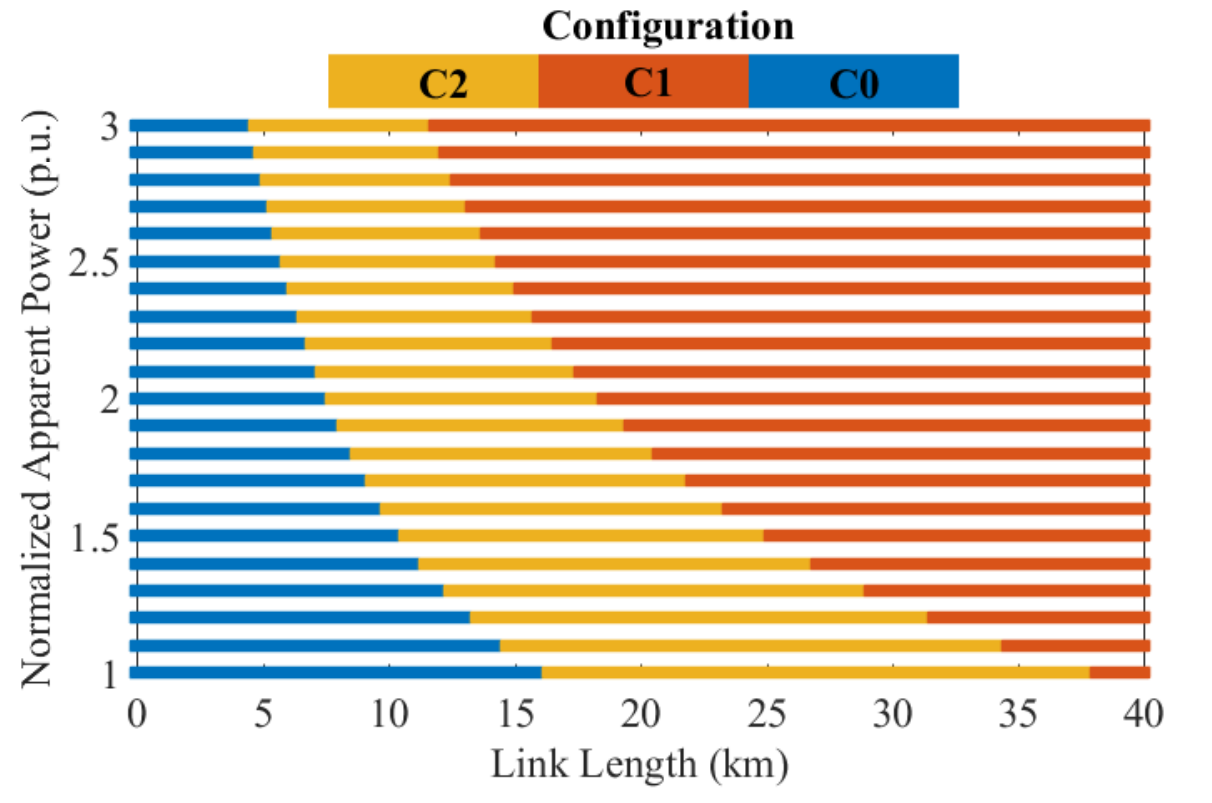}
   \caption{}
   \label{figbestC09} 
\end{subfigure}
\begin{subfigure}[b]{1\columnwidth}
   \includegraphics[width=1\linewidth]{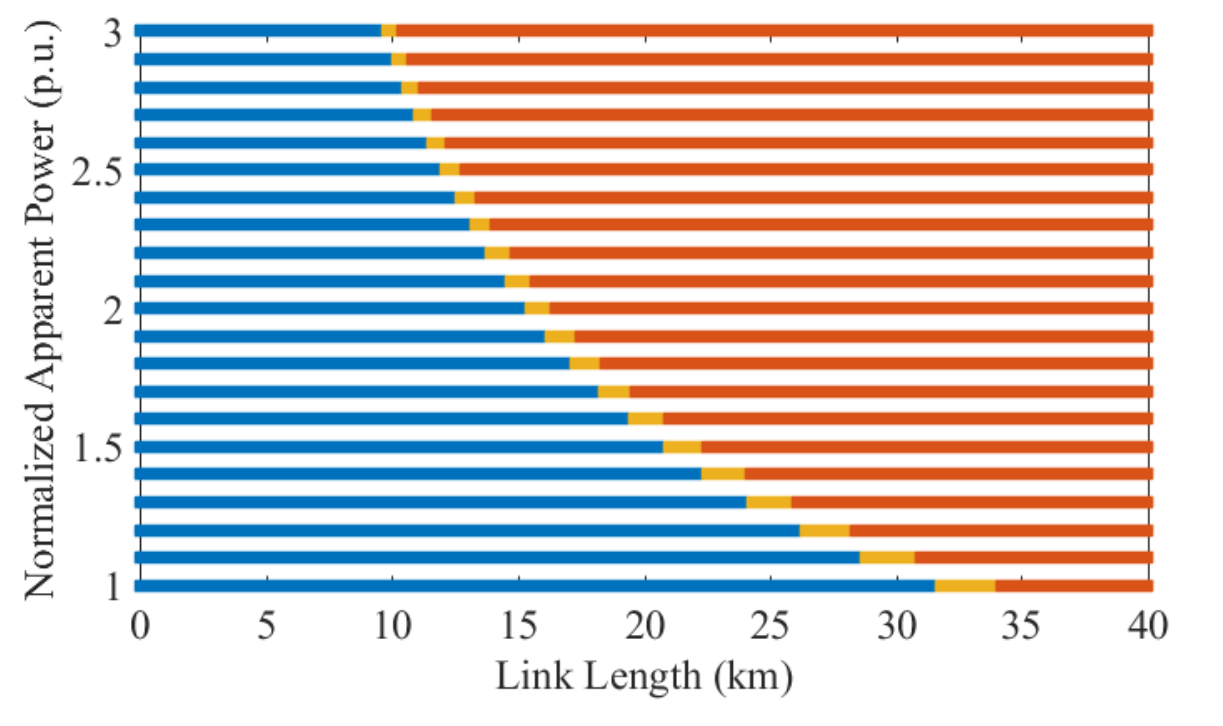}
   \caption{}
   \label{figbestC1}
\end{subfigure}
\caption[test]{Most Efficient operational configuration corresponding to link length and RSS power demand at (a) pf=0.9. (b)pf=1.} \label{figbestC}
\end{figure}

The minimum and maximum attainable crossover points $L_{\text{cr,A,min}}$ and $L_{\text{cr,C,max}}$, for given $S_{\text{actual}}$ and pf, can be determined for the optimal y from~Fig.~\ref{figlcross_C0C2} and~Fig.~\ref{figlcross_C1C2} respectively. Depending on the link length $L_{\text{link}}$, the configuration $C_{\text{n}}$ giving the maximum operational efficiency can be selected based on~\eqref{eqeboundaries},
\begin{equation}
C_{\text{n}}=
    \begin{cases}
      C_{\text{0}}, & \text{if}\ (L_{\text{link}}< L_{\text{cr,A,min}})\land (L_{\text{link}}< L_{\text{cr,B}}) \\
      C_{\text{1}}, & \text{if}\ (L_{\text{link}}\geq L_{\text{cr,B}})\land (L_{\text{link}}\geq L_{\text{cr,C,max}}) \\
      C_{\text{2}}, & \text{if}\ (L_{\text{link}}\geq L_{\text{cr,A,min}})\land (L_{\text{link}}< L_{\text{cr,C,max}}) 
    \end{cases}\label{eqeboundaries} \\
  \end{equation}
The maximum efficiency boundaries for selecting the operational configuration specific to the normalized RSS demand $S_{\text{actual}}$/$S_{\text{link}}$, pf and link length between SSS and RSS are given in~Fig.~\ref{figbestC}. From~Fig.~\ref{figbestC09}, it can be observed that as link length increases, configurations involving dc link operation become more efficient. Link length at which shift from pure ac configuration (C0, blue) to a parallel ac-dc operation (C2, yellow) occurs decreases as the power demand increases. This is because the conduction losses increase with square of the link current, while the converter losses increase linearly.  Similar trend with demand is observed in the shift from C2 (yellow) to pure dc operation (C1, red), but the rate at which this occurs is lower than the trend observed with C0 (blue) to C2 (yellow) shift. This is because the converter losses occur with both C1 and C2, albeit in different proportions. The inference from this behaviour is that the range of link length for which parallel ac-dc operation is the most efficient configuration decreases as the demand increases. For example, C2 is the most efficient configuration if the system link length is between 7.7\,km to 18.5\,km (range of 10.8\,km) with RSS demand at 2\,p.u at 0.9 pf. On the other hand, it reduces to that between 4.6\,km to 11.7\,km (range of 7.1\,km) with RSS demand at 3\,p.u at 0.9 pf. 

Since the C0-C2 crossover is observed to occur before that of C0-C1, parallel ac-dc operation can be used to extend the range in which ac to dc refurbishment configuration can be employed for capacity enhancement, while keeping the system efficiency maximum. For instance, at 3\,p.u RSS demand with 0.9 pf, C0-C2 crossover occurs at 4.6\,km, while it is 5.4\,km for C0-C1 crossover, implying that with C2 efficiency gains can be achieved with lower link lengths as compared to C1.

It can be observed in~Fig.~\ref{figbestC1} that the range for highest efficiency of configuration (C2-yellow) has considerably reduced  as compared to Fig.~\ref{figbestC09}. This is because with higher power factor, the pure ac system (C0-blue) become more favourable for lower link length. On the other hand, pure dc (C1-red) becomes more efficient at higher link length. Consequently, the blue and red regions `eat into' the yellow range as the power factor increases.
\section{Sensitivity Analysis with Various Influencing Factors}
\label{secsense}
The efficiency boundary described in Fig.~\ref{figbestC} is specific to the line voltage of 10\,kV  with converter efficiency of 99.34\,\% and aluminum link conductors of 400\,mm$^2$ cross-sectional area. A sensitivity analysis is carried out in this section to describe the variation in preferred configuration as these factors change. The results are presented for maximum ($L_{\text{C2,max}}$) and minimum ($L_{\text{C2,min}}$) link lengths for which configuration C2 is favourable. Based on~\eqref{eqeboundaries}, these represent the $L_{\text{link}}$ corresponding to max($C_{\text{2}}$) and min($C_{\text{2}}$) respectively. In relation to Fig.~\ref{figbestC}, $L_{\text{C2,min}}$ represents the transition from configuration C0 (blue) to C2 (yellow), while $L_{\text{C2,max}}$ represents  the transition from configuration C2 (yellow) to C1 (red).
\subsection{AC substation voltage}
\label{secvll}
\begin{figure}[!h]
\centering
\includegraphics[width=1\columnwidth]{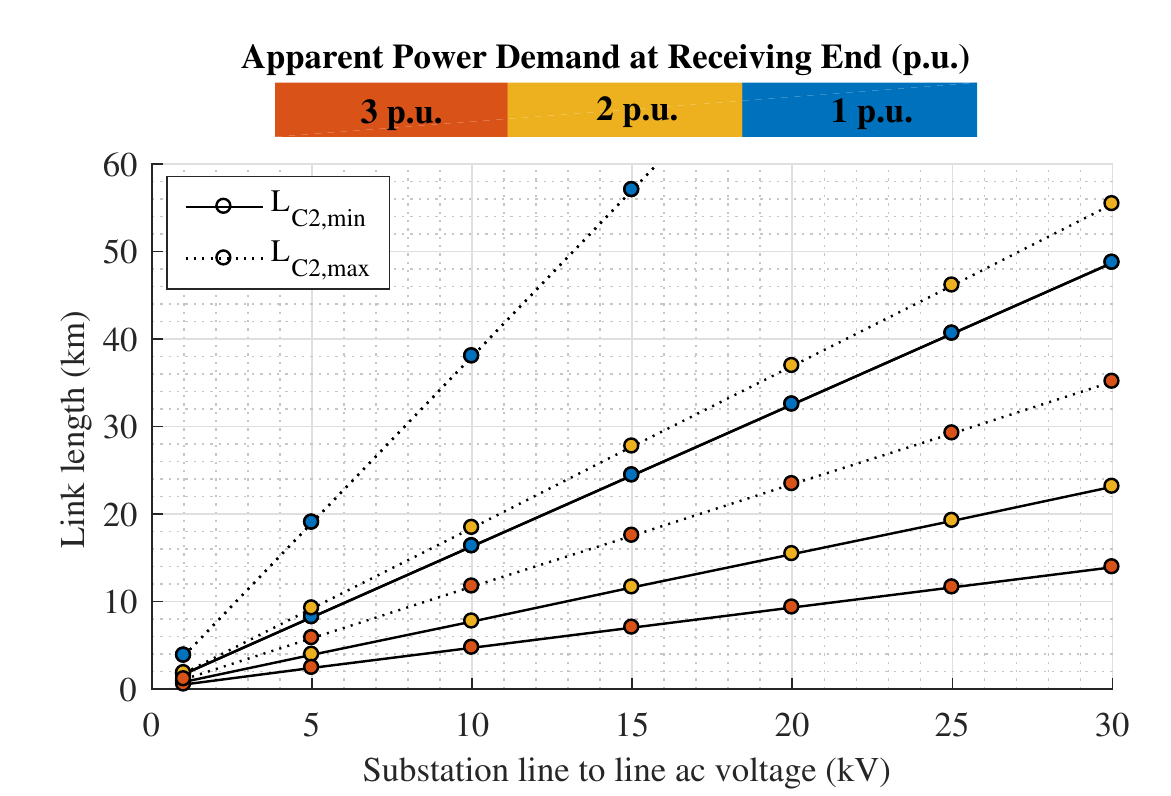}
\caption{Influence of SSS line to line ac voltage on the maximum and minimum link lengths for which C2 is the most efficient configuration for 400\,mm$^2$ link conductors delivering different normalized apparent power ($S_{\text{actual}}$/$S_{\text{link}}$) at 0.9 pf.}
\label{figvdep}
\end{figure}

Figure~\ref{figvdep} shows that $L_{\text{C2,max}}$ and $L_{\text{C2,min}}$ increase with $V_{\text{LL,rms}}$ for the same p.u. apparent power demand at RSS at 0.9 pf. This supports the intuitive understanding that as the ac voltage increases, the efficiency of the ac link operation improves, thus shifting the $L_{\text{link}}$ boundary further in the favour of ac based power delivery. This does not immediately imply a lower favor-ability for dc operation in practical scenarios because higher transmission voltages usually correspond to systems designed with longer link lengths. 

It is important to keep in mind that the base power used to normalize the MVA demand $S_{\text{actual}}$ is corresponding to the maximum MVA transfer capacity of individual ac or dc link ($S_{\text{link}}$) which increases linearly with $V_{\text{LL,rms}}$ for the same link conductor area. As a consequence, the direct square relationship between critical length and the substation voltage described by~\eqref{eqlcrit1}-\eqref{eqlcross} is translated to the linear dependence depicted in Fig.~\ref{figvdep}.

Another relevant observation is that the region encompassed by $L_{\text{C2,min}}$ and $L_{\text{C2,max}}$ for a particular apparent power demand widens with increasing substation voltage. This region defines the range of $L_{\text{link}}$ for which the parallel ac-dc configuration C2 is most efficient for the specified apparent power and ac voltage. The substation voltage dependent widening of the favoured C2 operation in Fig.~\ref{figvdep} increases as the p.u. apparent power demand decreases, which supports the observations made in~Fig.~\ref{figbestC}.
\subsection{Conductor Area}
\begin{figure}[!h]
\centering
\includegraphics[width=1\columnwidth]{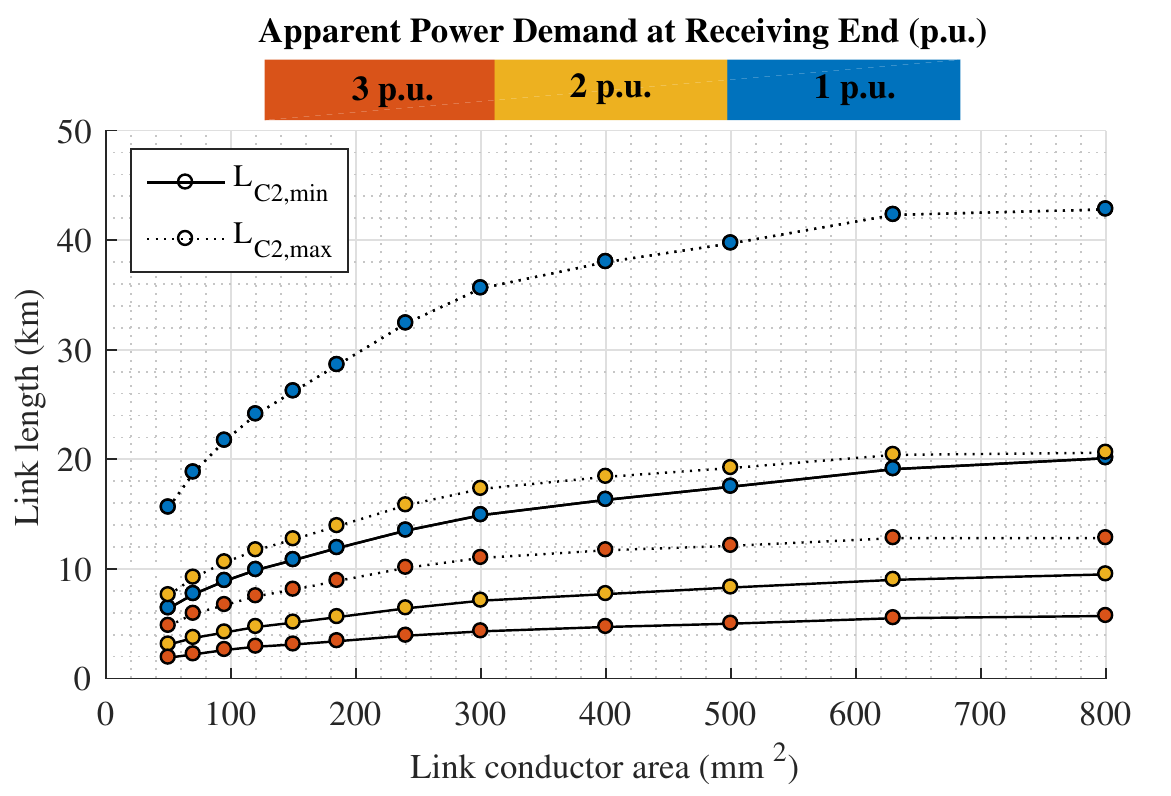}
\caption{Influence of link conductor cross-sectional area on the maximum and minimum link lengths for which C2 is the most efficient configuration at 0.9 pf for different normalized apparent power ($S_{\text{actual}}$/$S_{\text{link}}$) and $V_{\text{LL,rms}}=10$\,kV.}
\label{figrdep}
\end{figure}

It can be seen in Fig.~\ref{figrdep} that as the link conductor area increases,  $L_{\text{C2,min}}$ and $L_{\text{C2,max}}$ increase for specified p.u. RSS apparent power demand at 0.9\,pf and $V_{\text{LL,rms}}=10$\,kV. This is because with increasing area, the resistance of link length decreases and shifts the efficiency boundary in the favour of ac operation.

Again, the base power $S_{\text{link}}$ used to normalize the actual MVA demand $S_{\text{actual}}$ increases with cross-sectional area due to a corresponding increase in the rated current carrying capability of the conductor. The observations regarding widening of efficiency range for C2 has the same fundamental reasoning explained in Section~\ref{secvll}.
\subsection{Substation Converter Efficiency}
\begin{figure}[!h]
\centering
\includegraphics[width=1\columnwidth]{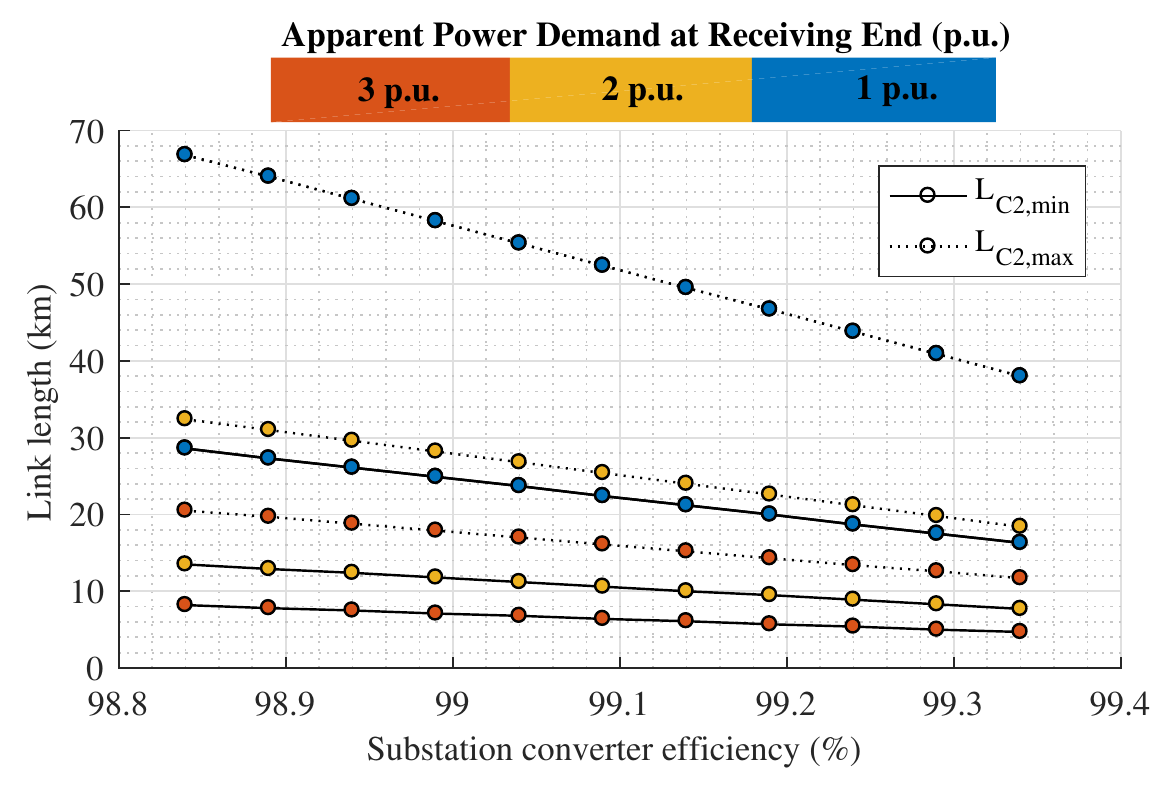}
\caption{Influence of the average converter efficiency on the maximum and minimum link lengths for which C2 is the most efficient configuration at 0.9 pf for different normalized apparent power ($S_{\text{actual}}$/$S_{\text{link}}$).}
\label{figeffdep}
\end{figure}

It can be observed in Fig.~\ref{figeffdep} that the $L_{\text{C2,min}}$ and $L_{\text{C2,max}}$ decrease with increasing average substation converter efficiency. The results are presented for $V_{\text{LL,rms}}=10$\,kV and A=400\,mm$^2$. The straightforward inference is that dc operation is favoured when converter efficiency improves, a behaviour that is mathematically presented in in~\eqref{eqlcrit1}-\eqref{eqlcross}. 

The region in which C2 is most efficient expands as the average converter efficiency decreases. This trend depends on the slope of $L_{\text{cr,A}}$, $L_{\text{cr,B}}$ and $L_{\text{cr,C}}$ described by ~\eqref{eqlcrit1}-\eqref{eqlcross} with respect to $\eta$ in relation to the conditions expressed in~\eqref{eqeboundaries}.
\section{Conclusions}
\label{secconc}
In this paper, different configurations using ac to dc refurbishment techniques were compared in terms of the operating efficiency for different power demands, power factors, dc link power share and link length with respect to the conventional ac capacity enhancement solution. All system configurations had the same power delivery capacity of 3\,p.u at 0.9 pf during (n-1) contingency.

The main finding is that a parallel ac-dc configuration can offer superior efficiency as compared to complete dc or complete ac system within a certain operating range. By determining the corresponding efficiency boundaries, it was shown that this range is 4.6-11.7\,km when a 10\,kV system is operating at its maximum design capacity of 3\,p.u at 0.9\,pf. Complete ac system has better efficiency for shorter link lengths, while complete dc operation is favourable for lengths above this range.

Sensitivity analysis shows that the range of link lengths for which parallel ac-dc distribution link configuration is the most efficient choice increases with decreasing apparent power demand and average converter efficiency or increasing substation ac bus voltage and link conductor areas. The corresponding minimum and maximum values of this range also increase. The feasibility of this configuration increases if the power factor at the receiving end is lower.

\bibliographystyle{ieeesty}
\bibliography{lossana}

\end{document}